\documentstyle[12pt,epsfig]{article}

\begin{document}   

\def\beq{\begin{equation}}
\def\eeq{\end{equation}}
\def\eq{\beq\eeq}
\def\beqn{\begin{eqnarray}}
\def\eeqn{\end{eqnarray}}
\relax
\def\ba{\begin{array}}
\def\ea{\end{array}}
\def\squ{\tilde{Q}}
\def\tb{\mbox{tg$\beta$}}
\def\hH{\hat{H}}
\def\tu{\tilde{u}}
\def\td{\tilde{d}}
\def\te{\tilde{e}}
\def\tQ{\tilde{Q}}
\def\tL{\tilde{L}}
\def\tt{\tilde{t}}
\def\tb{\tilde{b}}
\def\ttau{\tilde{\tau}}
\def\tnu{\tilde{\nu}}
\def\ra{\rightarrow} 
\newcommand{\nn}{\nonumber}
\newcommand{\lsim}{\raisebox{-0.13cm}{~\shortstack{$<$ \\[-0.07cm] $\sim$}}~}
\newcommand{\gsim}{\raisebox{-0.13cm}{~\shortstack{$>$ \\[-0.07cm] $\sim$}}~}
\newcommand{\s}{\\ \vspace*{-2mm} }
\renewcommand{\theequation}{\thesection.\arabic{equation}}

\begin{flushright}
PM/98--16 \\
GDR--S--014\\
hep-ph/9807336
\end{flushright}

\vspace{1cm}

\begin{center}

{\large\sc {\bf 
Inverting the Supersymmetric Standard Model Spectrum: \\
\vspace{.5cm} 
from Physical to Lagrangian Ino
Parameters\\}} 
\vspace{1cm}
{ J.-L. Kneur~\footnote{kneur@lpm.univ-montp2.fr} and 
G. Moultaka~\footnote{moultaka@lpm.univ-montp2.fr}}
\vspace{.5cm}

Physique Math\'ematique et Th\'eorique, UMR No 5825--CNRS, \\
Universit\'e Montpellier II, F--34095 Montpellier Cedex 5, France.

\end{center}

\vspace*{1.5cm}

\begin{abstract}
\setlength{\baselineskip}{20pt}

\noindent 
We examine the possibility of recovering  
the supersymmetric (and soft supersymmetry breaking) Lagrangian 
parameters as direct {\em analytical} expressions of appropriate
physical masses, for the unconstrained (but CP and
R-parity conserving) minimal supersymmetric standard model. 
We concentrate 
mainly on the algebraically 
non--trivial ``inversion" 
for the ino 
parameters, and obtain, for given values of $\tan\beta$,
 simple analytical expressions for the
$\mu$, $M_1$ and $M_2$ parameters in terms of  
three arbitrary
input physical masses, namely either
two chargino and
one neutralino masses, or alternatively 
one chargino and two neutralino masses.   
We illustrate and discuss in detail the possible occurrence of 
ambiguities in  this reconstruction.
The dependence of the resulting ino Lagrangian parameters upon physical masses
is illustrated, and some simple generic behaviour  
uncovered in this way.
We finally briefly sketch generalizing such an
inversion to the full set of MSSM Lagrangian parameters.
\end{abstract}

\newpage
\setlength{\baselineskip}{15pt}

\section{Introduction}
\setcounter{footnote}{0}
In the
Minimal Supersymmetric extension 
of the Standard Model (MSSM)~\cite{R1, martin}, 
without additional theoretical assumptions, the necessary breakdown
of supersymmetry leads to a 
large number of arbitrary parameters, consisting of 
all possible renormalizable soft supersymmetry breaking
terms~\cite{softbreak}. 
The latter arbitrariness simply reflects 
our present ignorance of the fundamental mechanism underlying 
the breakdown of supersymmetry.
Although several theoretically appealing scenarios have been explored,
with additional assumptions considerably reducing 
the arbitrariness of the soft SUSY-breaking
sector (typically like in hidden sector supergravity models~\cite{sugra},
or gauge-mediated susy breaking models~\cite{gaugemed}), 
alternative strategies may be 
useful to exploit the information from the present and future
collider data. If SUSY is realized
at low energy in some way, and some super-partners
discovered, the next immediate 
task would be to
reconstruct from the data the structure
of the SUSY and soft-SUSY breaking 
Lagrangian [that is, to determine 
as precisely as possible how the experimentally measured 
parameters would translate as possible values (or bounds) on the
MSSM Lagrangian parameters.] \\
The expressions of physical parameters (mass eigenvalues, 
mixing angles and physical couplings) as functions of the most general, 
unconstrained MSSM Lagrangian, are well known~\cite{R1,R2}. Now, 
inverting those relations (i.e. expressing the Lagrangian
parameters in terms of physical parameters) is non trivial 
in the unconstrained MSSM, due to the large number of parameters
and possible mixing among them. This is especially true 
in the ino sector which involves a  
relatively complicated structure of the mixing terms, 
with e.g. a 4 x 4 mass matrix to ``de--diagonalize" in the
neutralino sector. To our knowledge, no systematic 
analytical inversion has been explored up to now.\\

The aim of the present paper is twofold: 
\begin{itemize}
\item
Investigate as much as possible
analytically the reconstruction of the Lagrangian parameters from the physical
masses, especially in the ino sector, identifying clearly the procedure
and the related ambiguities when they occur;
\item construct a numerical code based on algorithms using as much as 
possible the above mentioned results, and which can be readily used for a full 
numerical study of the Lagrangian parameters as a function of the physical 
masses. 
\end{itemize}

For definiteness we will restrict ourselves throughout the paper to the (CP and
R-parity conserving) unconstrained MSSM. Also to simplify the presentation
we will consider two illustrative ``scenarios'' where either two chargino
and one neutralino masses, or one chargino and two neutralino masses 
are input. The outcome is a well--defined algorithm
providing, for given $\tan\beta$,
the values of the $\mu$, $M_1$ and $M_2$ parameters in terms of  
three arbitrary input masses. 
Furthermore, one of our main results
is the derivation, in terms of simple analytic expressions, 
of the full neutralino mass spectrum and one
of the Lagrangian parameters ($M_1$), when the three other parameters  
($M_2, \mu, \tan \beta$) and one neutralino mass are given. This together
with the analytical expressions of the chargino sector constitute the
building blocks of our algorithms.\\
  
It should be stressed here that we are not aiming, in this paper,
at a specific phenomenological reconstruction of the MSSM parameters from
experimental observables. Clearly 
an algebraic approach 
cannot replace more systematic phenomenological studies of the 
reconstruction of the basic MSSM parameters. 
%Different envisaged 
%experimental strategies to extract
%the MSSM physical spectrum from present and future collider data,
Various experimental strategies to extract the MSSM physical spectrum from 
present and future collider data have been considered, together with a detailed 
assessment on mass measurements, e.g., in ref.~\cite{SNOWMASS,Denegri}.
%and in particular the assessment on  mass measurements have been studied
%in detail, 
%e.g., in ref.~\cite{SNOWMASS,Denegri}.
On the other hand, a complete scanning of the unconstrained MSSM parameters
would be in practice rather tedious and extremely
time consuming. 
We believe that our study would be useful 
and rather flexible  for
such reconstructions: besides the fact 
of providing a fast numerical inversion routine, 
we systematically encode the possible ambiguities that can arise through the 
reconstruction from some of the physical masses. 
In the most optimistic case (that is, if knowing a sufficient 
number of  physical input ino masses
and with sufficient accuracy),
this inversion
allows for a precise reconstruction of the
unconstrained MSSM ino sector Lagrangian. If a
more limited experimental information of the ino spectrum were
available, such as only mass bounds or mass differences typically, 
one could still derive corresponding constraints 
on the MSSM Lagrangian parameters as well.\\
%Note also that it 
%is more or less possible, at least in principle, to use 
%the direct ``standard" approach to obtain 
%the requested information back on the basic parameters, 
%e.g. by some 
%systematic ``scanning" of 
%the whole Lagrangian parameter space (this is indeed 
%what is partly used 
%in some past or recent simulations\cite{SNOWMASS, Denegri}).
%But as we already stressed, it should be advantageous
%to have a relatively simple (and thus fast) 
%inversion algorithm,  since in 
%practice a complete scanning of the unconstrained MSSM parameters
%would be rather tedious, extremely
%time consuming, and probably not particularly illuminating. \\
 
 Our input/output parameters are 
deliberately chosen so as to render the analytical inversion 
the most simple and transparent.  
As will be 
explained in details later on, our algorithms will be either fully
analytical, in which case at most a readily solvable {\sl cubic} equation
is involved (in contrast with
the generically quartic one in the ordinary diagonalization of the
neutralino sector~\cite{egyptiens}),
or needs some numerical iteration, in which case it
relies essentially on a system of two algebraic 
equations, that are only {\em quadratic} (in  
contrast with a ``brute
force" inversion, which would involve a highly non--linear
set of equations). It thus 
demonstrates the feasibility and relative simplicity
of such a systematic inversion and more importantly we also 
exhibit the general trend 
and sensitivity of the Lagrangian parameters to the physical masses. 
Note, however, that we
do not assume at this level any knowledge of the couplings of inos to higgses
and gauge bosons. These would of course lead to extra information \cite{choi}
which together with our results could allow an even  more 
complete reconstruction and cross-checks for the
ino sector. \\

The rest of the paper is organized as follows. 
In section 2, we  briefly recall the main content of the MSSM
Lagrangian, just to fix our definitions and 
conventions. 
In section 3, we present our main 
results for the analytic
inversion of the ino parameters, 
and discuss uniqueness conditions. Some essentially technical
material is given in two appendices.
In section 4 we illustrate the dependence
of the ino sector Lagrangian
 parameters, as  functions of the relevant
physical input masses. We also illustrate
some resulting values of the ino Lagrangian 
parameters at a grand
unification (GUT) scale.
In section 5 we briefly outline, for completeness, similar inversion 
relations in the scalar fermion and Higgs sector of the MSSM.
Finally section 6 presents the conclusions and an outlook.     

\section{MSSM Lagrangian and spectrum: reminder}
\setcounter{equation}{0}
The supersymmetric part of the unconstrained MSSM Lagrangian 
consists of the gauged kinetic terms for the
$SU(3)\times SU(2)\times U(1)$ vector multiplets,  involving the three
gauge couplings $g_3$, $g_2$, and $g_1$, respectively, 
the gauge invariant kinetic term of matter fields, and
the superpotential:
\beq
\label{superW}
W = \hat{\bar{u}} Y_t \hat{Q} \hH_u - \hat{\bar{d}} Y_b \hat{Q} \hH_d  
     -\hat{\bar{e}} Y_\tau \hat{L} \hH_d 
+\mu \hH_u \hH_d \;.
\eeq
In eq.~(\ref{superW}), the hats 
denote superfields, $Y_i$ are the Yukawa couplings,
and $\hH_u \hH_d \equiv \epsilon_{ij} \hH^i_u \hH^j_d$,
(with $\epsilon_{12}=1$ ) 
fixes our sign convention for $\mu$.  
In (\ref{superW}) and in the following 
we will in fact neglect all 
flavor non-diagonal terms. We also 
suppress any flavor, color, etc indices. 
Moreover, we will restrict here to the case where all 
parameters are assumed to be  real. \\
 
The soft SUSY-breaking terms  involve \\
--the trilinear coupling terms:
\beq
\label{Lsofttri}
{\cal L}_{trilinear} = -(\tu A_t Y_t H_u \tQ +\td A_b Y_b H_d \tQ 
+\te A_\tau Y_\tau H_d \tL +h.c.)\;;
\eeq
involving the $A_i$ parameters (which have a mass dimension 
in this convention); \\  
--a contribution to the sfermion mass terms: 
\beqn
\label{Lsfermion}
{\cal L}_{sfermion} = -M^2_Q (\tt^\star_L \tt_L +\tb^\star_L \tb_L)
-M^2_{t_R} \tt^\star_R \tt_R - M^2_{b_R} \tb^\star_R \tb_R \\
\nn
-M^2_L (\ttau^\star_L \ttau_L +\tnu^\star_L \tnu_L) -M^2_{\tau_R}
\ttau^\star_R \ttau_R
\eeqn
and the gaugino mass terms: \\
\beq
\label{Lgaugino}
{\cal L}_{gaugino} = -\frac{M_1}{2} \tilde{B} \tilde{B} 
-\frac{M_2}{2} \tilde{W}^i \tilde{W}_i
-\frac{M_3}{2} \tilde{G}^a \tilde{G}_a \;;
\eeq
which fixes our sign conventions.
There are also supersymmetric contributions to the ino masses, of which
we write here for illustration only the ones contributing to neutralinos
(after electroweak symmetry breaking): 

\begin{eqnarray}
\label{Lneutralino}
{\cal L}_{neutralino}&=&  
 m_Z c_w \sin \beta \tilde{W}_3 \tilde{H}_u
- m_Z c_w \cos \beta \tilde{W}_3 \tilde{H}_d \nonumber \\
&& + m_Z s_w \cos \beta \tilde{B} \tilde{H}_d 
   - m_Z s_w \sin \beta \tilde{B} \tilde{H}_u +\mu \tilde{H}_u \tilde{H}_d
\nonumber \\ 
\end{eqnarray}
with $s_w \equiv \sin\theta_W$, $c_w \equiv \cos\theta_W$, and
$\tan\beta \equiv v_u/v_d$ the ratio of the two Higgs vacuum
expectation values, $v_{u,d} = \langle H_{u,d} \rangle$. 
It is now easy to see from Eqs.(\ref{Lgaugino}, \ref{Lneutralino}) that
only the relative phases of say $(M_2, M_1)$ and $(M_2 , \mu)$ are physically
relevant.  
Indeed any phase change of $M_2$ in Eq.(\ref{Lgaugino}) can always be
absorbed by a phase change of the $\tilde{W}$ field. The latter however
fixes uniquely the phase change of $\tilde{H}_d$, $\tilde{H}_u$
and $\tilde{B}$ in Eq.(\ref{Lneutralino}) in such a way that
the phases of the combinations $M_2/M_1$ and $\mu M_2^2 $ remain unchanged. 
The discussion is of course  
reminiscent of the one carried out for instance in ~\cite{martin}. 
In the present study, as already mentioned, we restrict ourselves to real-valued parameters, but do not necessarily assume universality of gaugino masses. 
 We can thus choose, without loss of generality, 
$M_2$ to be {\em always positive} and $M_1$ and $\mu$ to have arbitrary 
signs. \\  

Finally, 
the Higgs potential is built from soft SUSY breaking and F-term
contributions to the Higgs scalar mass terms 
plus quadrilinear D-terms, and reads, before
$SU(2)\times U(1)$ breaking:
\beqn
\label{VHiggs}
V_{Higgs} = (m^2_{H_d} +\mu^2) \vert H_d \vert^2 
+(m^2_{H_u} +\mu^2) \vert H_u \vert^2 +(B \mu H_u H_d + h.c.) \\
\nn
+\frac{g^2_1+g^2_2}{8} (\vert H_d \vert^2 -\vert H_u \vert^2 )^2
+\frac{g^2_2}{2} \vert H^{\star i}_d H_{u,i} \vert^2 +V^{1-loop}_{eff} 
\eeqn
which also fixes a sign convention for the $B$ parameter,
and where  $V^{1-loop}_{eff}$
is the one-loop contribution to the effective potential~\cite{Vloop,rge,HiggsRC, others}.  
The corresponding 
Higgs masses and mixing angles (after electroweak symmetry breaking)
are given in Appendix 1.

\section{Inverting the ino MSSM spectrum}
\setcounter{equation}{0}
Let us first sketch our general procedure to reconstruct the
ino sector parameters. As already mentioned in the 
introduction, we have to fix a 
specific choice of input/output parameters, 
that we do as follows. We first assume that $\tan\beta$ is an input parameter
 at this stage, i.e. that it has been extracted from elsewhere 
prior to ino reconstruction, or simply fixed arbitrarily. (Obviously, 
once a reconstruction algorithm is defined for fixed $\tan\beta$, 
one may always study the sensitivity to this parameter). 
Then, we consider two basic scenarios, $S_1$ and $S_2$, and the 
corresponding algorithms.
\begin{itemize} 
\item[$S_1$:] the input are {\em two} chargino masses, $M_{\chi^+_1}$,
$M_{\chi^+_2}$, and one
(but any) of the neutralino masses, $\pm M_{N_i}$. 
\item[$S_2$:] the input are 
 a single chargino mass, $M_{\chi^+_1}$, and two neutralino masses, say
$\pm M_{N_2}$, $\pm M_{N_3}$.
\end{itemize}
It is important to note here that we will adopt the formulation in which
the eigenvalues of the chargino mass matrix are by construction always
positive, while those of the neutralino mass matrix can have either signs.
This means that in the inversion process one should consider both
$\pm M_{N_i}$ as possible inputs [a feature consistently taken into account
in the formulae we derive.]

For each of the above scenarios the aim will be to identify a 
corresponding algorithm which allows the determination of the output values
of $\mu$, $M_2$, $M_1$, as well as the remaining three neutralino
masses (resp. two neutralinos and one chargino masses) in the $S_1$
(resp. $S_2$) case.\\

 First, as we will show in detail below and in 
Appendix \ref{app2}, scenario $S_1$ allows a fully analytical algorithm  to
handle all the features of the inversion procedure.
Algorithm $S_1$ will therefore be our starting point for the whole study
\footnote{Except when ambiguous, we call from now on the scenario 
and its associated algorithm with the same name.}. We stress, however,
that it does not 
necessarily imply a strong 
prejudice on the ``chronology" of discovery of the inos since, as we
shall explain, it may also be used as a basic building block for a 
probably more
phenomenologically relevant situation, in scenario $S_2$ below.
[Furthermore, we will see that $S_1$ can be naturally separated into two 
independent steps 
corresponding to chargino and neutralino sectors. It can thus also be
readily used in a more general context than the one of the present paper].\\
  
To scenario $S_2$ we associate the following algorithm:
first assuming {\em temporarily} that e.g. $\mu$  
is an input parameter, together with two 
neutralino masses, say $M_{N_2}$ and $M_{N_3}$, 
then a simple quadratic system 
gives $M_2$, $M_1$ as function of the latter input, 
as well as all other ino masses. 
Now, the 
key observation is that it is relatively simple to merge
the solutions of the latter system with the previous algorithm $S_1$,  to obtain 
consistently $\mu$, $M_2$, $M_1$ from  a single chargino mass,
$M_{\chi^+_1}$, and two neutralino masses, say
$M_{N_2}$, $M_{N_3}$ input. More precisely, choosing an adequate 
``initial guess" value for e.g. $M_2$
one can simply use $S_1$ (actually only 
a part
of it) to determine $\mu$ in terms of ($M_{\chi^+_1}$, $M_2$), followed by 
the above mentioned $M_1$, $M_2$ solutions, iterating with respect to
$M_2$ until a convergent
set of values is obtained~\footnote{There are always (complex) solutions
since, as easily checked, the  explicit 
non-linear system equivalent to $S_2$ 
is a (high degree) polynomial. Non convergent 
domains thus simply correspond
to the impossibility to match our  
basic assumption of real 
$\mu, M_1, M_2$.}. 
While $S_1$ was fully analytical,
the price to pay is that, accordingly, $S_2$ has 
to be partly numerical through
the iterative procedure.
In most cases, this 
combined algorithm turns out to converge rapidly, 
after 2 to 3 iterations for an accuracy that is sufficient for all practical
purposes.\\

This particular decomposition, with this choice of 
input/output, is the one giving the most algebraically tractable
inversion in the
ino sector. 
In fact, while $S_1$ alone 
is probably not very relevant physically (since 
for a rather generic choice of the Lagrangian parameters, it looks more
likely to have two neutralinos and one chargino in the lightest part
of the ino spectrum~\cite{SNOWMASS,martin,others}), precisely
this situation is tractable from  the combined algorithm of scenario $S_2$, 
as explained above. However, it is rather instructive to
study algorithm $S_1$ separately in some detail, as it  
exhibits important properties
of the inversion in a relatively simple manner.
[Of course it is also possible to use any part of our particular 
procedures to simply scan over some values, if not known, of some
of the ino masses, as will be 
illustrated in the plots in section 4]. 
     
\subsection{Basic algorithm $S_1$: $M_{\chi^+_1}$,
$M_{\chi^+_2}$, $M_{N_2}$ input}
 Assuming $\tan\beta$
and the two chargino masses given, one easily obtains from
the expressions of the chargino mass eigenvalues (see eq~(\ref{Mchi12}) in
Appendix \ref{app1}):  
\beq
\label{invmu2}
\mu^2  = \frac{1}{2}(M^2_{\chi^+_1}+M^2_{\chi^+_2}-2m^2_W 
\pm [ (M^2_{\chi^+_1}+M^2_{\chi^+_2}-2m^2_W)^2 -4(m^2_W \sin 2\beta \pm 
M_{\chi^+_1} M_{\chi^+_2})^2 ]^{1/2} )
\eeq
and
\beq
\label{invM2}
M_2  =  [M^2_{\chi^+_1}+M^2_{\chi^+_2}-2m^2_W -\mu^2 ]^{1/2} 
\eeq 
In eqs.~(\ref{invmu2}) and (\ref{invM2})
the $\pm$ ambiguity in front of the square root results from
the invariance of the
physical masses (see eq.~(\ref{Mchi12}) of appendix A) 
under the substitution $\vert \mu\vert \leftrightarrow M_2$
 (we stress again 
our sign convention, $M_2 >0 $ in (\ref{invM2})). 
In other words, 
from the two chargino masses input only, there is clearly no way of 
assessing the amount of gaugino or higgsinos components of each of them.
To achieve this, one would require the knowledge of mixing angles
(i.e. couplings)~\cite{choi} or alternatively of two neutralino masses,
as will be discussed below in scenario $S_2$.
In what follows, we thus arbitrarily 
{\em choose} to illustrate only the case
$\vert\mu\vert \leq M_2$ (corresponding to the minus sign choice
in front of the square root of (\ref{invmu2})), i.e. the case where
the lightest chargino has a stronger higgsino like component. We thus
refer to this situation as ``higgsino-like''.
The resulting output of
algorithm $S_1$ corresponding to the opposite ``gaugino-like'' situation, 
$\vert\mu\vert \geq M_2$, is therefore 
trivially obtained by interchanging
the values of $\vert \mu\vert $ and $M_2$ (together with the correct
sign of $\mu$ assignment, see e.g.
captions of fig.~\ref{fig2}).   
In addition, since $M_2 >0 $, $M_1$ 
will have arbitrary sign (see previous section 2).
[Had we chosen to let the sign of $M_2$ rather than $M_1$ 
undetermined at this stage, 
our algorithm would
have been slightly more complicated].\\  
Concerning now the sign ambiguity on $\mu$, 
it is in fact fixed from
another relation from eqs.~(\ref{Mchi12}) (implicitly
used in eq.~(\ref{invmu2})):
\beq
\label{sgnmu}
M_2 \;\mu = m^2_W \sin 2\beta \pm 
M_{\chi^+_1} M_{\chi^+_2}\;
\eeq
for each respective choice of the $\pm $ ambiguity. The latter $\pm$,
also appearing inside the square root in (\ref{invmu2}), 
corresponds to a true 
ambiguity, i.e. whenever the expression under 
the square root of eq.~(\ref{invmu2}) is positive for 
both sign choice, there are two distinct 
solutions for ($\mu$, $M_2$). 
Obviously, the occurrence of this discrete twofold ambiguity 
depends crucially  on the values of $M_{\chi^+_1}$, 
$M_{\chi^+_2}$ and $\tan\beta$, and deserves a more careful examination
to which we now turn.\\

\begin{figure}[htb]
\epsfxsize=145mm
\epsfysize=120mm
\begin{center}
\hspace*{0in}
\epsffile{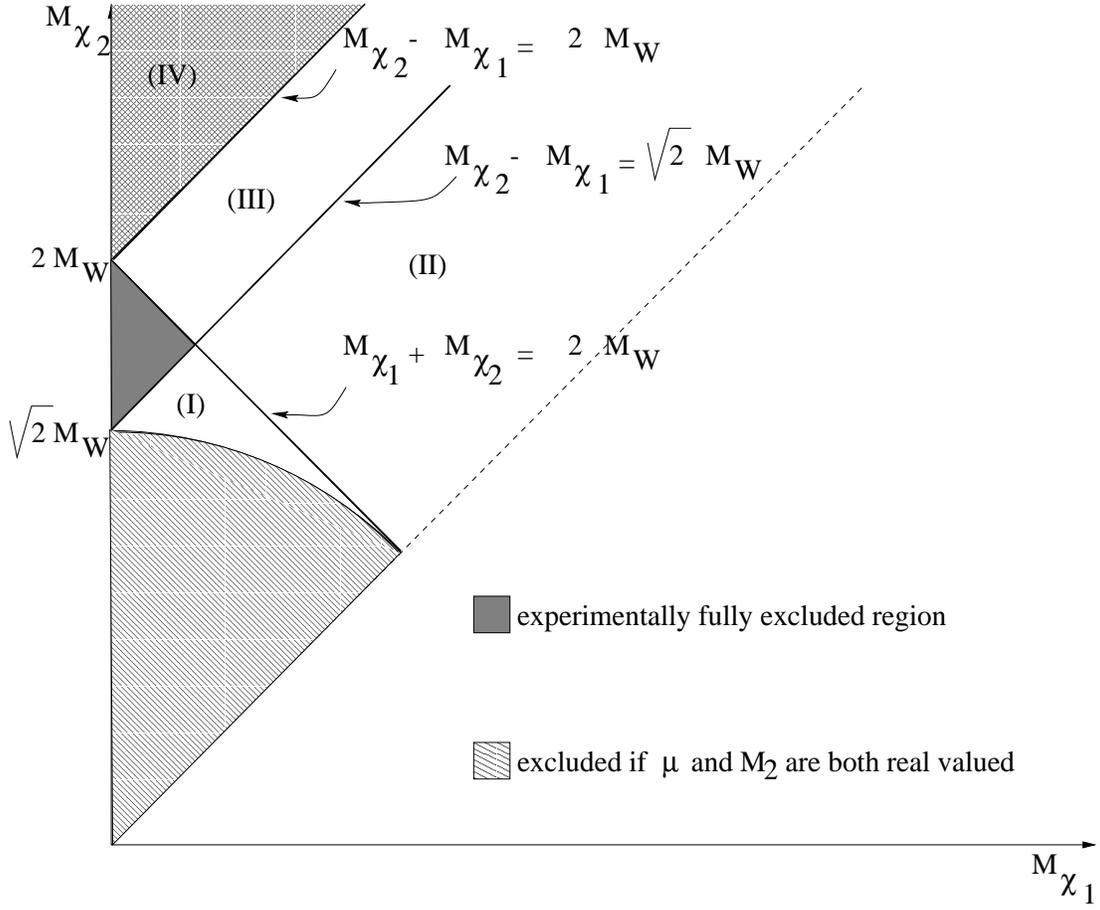}
\vspace*{0.1in}
\caption{ \label{fig1} Domains of solutions
for  $\mu$, $M_2$ as functions of  
$M_{\chi^+_1}$, $M_{\chi^+_2}$ and $\tan\beta$, where we assumed without
loss of generality $M_{\chi^+_2} > M_{\chi^+_1}$. The
detailed definitions and properties of the different domains
(I)--(IV) are explained in Table 1.}
\vspace*{-0.1in}
\end{center}
\end{figure}

We illustrate in table 1 and 
fig.\ref{fig1} 
the different domains where one obtains either no solutions 
(for $\mu$ real), a unique solution, or the discrete 
twofold ambiguity,
depending on 
the input values $M_{\chi^+_1}$, $M_{\chi^+_2}$ 
and $\tan\beta$. What is relevant here
are the chargino mass difference
 $\Delta_{\chi^+} \equiv  |M_{\chi^+_2}-M_{\chi^+_1}|$
and sum 
$\Sigma_{\chi+} \equiv M_{\chi^+_1} + M_{\chi^+_2}$, as well as
the quantity
\begin{equation}
X^{\epsilon_1}_{\epsilon_2} = - 
\frac{ \epsilon_1 2 M_{\chi^+_1} M_{\chi^+_2} + \epsilon_2 ( M_{\chi^+_1}^2 +
M_{\chi^+_2}^2 - 2 M_W^2)}{2 M_W^2}\;,
\end{equation}
where $ \epsilon_1, \epsilon_2 = \pm 1$.
For instance, the last two columns of table 1 summarize respectively 
the occurrence of the twofold ambiguity or the existence of (real) solutions
at all. In some zones this is fully controlled by the values of $\tan \beta$
which we expressed in the form of necessary and sufficient constraints
on $\sin 2 \beta$.    
This is systematically taken into account in our algorithm,
which gives in the relevant case the two possible solutions for $\mu$. 
One then needs 
an extra information to eventually eliminate one of
the solutions. Note finally that in fig.\ref{fig1} we assumed for
simplicity that $M_{\chi^+_2} > M_{\chi^+_1}$ which is obviously just a matter
of convention. [In any case, the opposite situation corresponds simply to
a symmetry around the bisectrix line.] 

\begin{table}[htb]
%\center{
\begin{tabular}{|c|c|c|l|l|l|}
\hline\hline
\hspace*{-0.2cm}$\Delta_{\chi^+}\! /\!\sqrt{2} M_W$
         & \hspace*{-0.2cm}$\Delta_{\chi^+} \! / 2 M_W $& 
$ \!\Sigma_{\chi+} \!/ 2 M_W$ & $\!X_{+}^{+} \!<\!0 ; X_{+}^{-} \!<\!1$   
& \hspace*{-0.2cm}twofold  &constraint  \\ 
    &     &     &  in all cases                 &\hspace*{-0.2cm}ambiguity 
& \hspace*{-0.2cm} on $\beta$\\ \hline
$< 1$ & $< 1$ & $< 1$ &\hspace*{-0.2cm}  $0 \!<\! X_{\pm}^{-}\!<\! 1$    
      &\hspace*{-0.2cm} No & \hspace*{-0.2cm}
                                     $ X_{+}^{-} \!< \! \sin 2 \beta \!< \!
 X_{-}^{-}$\\ 
 zone (I) &  &    &  \hspace*{-0.2cm}$X_{-}^{+} \!< \!0 $               &  &  
 \\ \hline
$< 1$ & $< 1$ & $> 1$ &\hspace*{-0.2cm} $0 \!<\! X_{+}^{-} \!<\!1 \!<\! 
X_{-}^{-}$ &\hspace*{-0.2cm} No & 
                                     $ \sin 2 \beta > X_{+}^{-}$\\ 
 zone (II) &  &  &\hspace*{-0.2cm}  $X_{-}^{+} \!<\!0 $                & & \\ 
\hline
$> 1$ & $< 1$ & $< 1$ &\hspace*{-0.2cm}  exp. excluded     
            & & \\ \hline
$> 1$ & $< 1$ & $> 1$ &\hspace*{-0.2cm}  $X_{+}^{-} \!<\! 0 ; 1 \!<\! X_{-}^{-} $ &\hspace*{-0.2cm} Yes when 
&\hspace*{-0.2cm} No\\ 
 zone (III) &   &     & \hspace*{-0.2cm} $0\!<\! X_{-}^{+} \!<\!1$                
                                 &\hspace*{-0.2cm} $\sin 2 \beta\!<\! X_{-}^{+}$ & \\ \hline
$> 1$ & $> 1$ & $> 1$ &\hspace*{-0.2cm} $ X_{+}^{-} \!<\! 0 ; 1\!<\! X_{-}^{-}$ &\hspace*{-0.2cm} Yes &\hspace*{-0.2cm} No \\
 zone (IV) &  &  &\hspace*{-0.2cm}  $X_{-}^+ >1 $                 & & \\ \hline 
\end{tabular}
%}
\caption{Inequalities characterizing the different domains of solutions
$\mu$, $M_2$ for arbitrary $M_{\chi^+_1}$,
$M_{\chi^+_2}$ input.  
$\Delta_{\chi^+} \equiv  \vert M_{\chi^+_2}-M_{\chi^+_1}\vert $;
$\Sigma_{\chi+} \equiv M_{\chi^+_1} + M_{\chi^+_2}$.  
$X^{\pm}_{\pm}$ is defined in the text.}
\end{table}

\subsection{Neutralino mass inversion}
Let us now turn to the de-diagonalization of the neutralino sector.
The question we want to answer analytically here, is how to determine $M_1$ 
and three neutralino masses, when  $M_2$, $\mu$, $\tan \beta$ and a fourth 
neutralino mass are given.
We will only outline the procedure leaving some technical details
to the appendix. As we shall see, the discussion in this section
is quite general and serves indifferently as a basis for the more specific
algorithm $S_1$ or $S_2$.\\
 
Since we restrict ourselves to the case where $M_1$ , $M_2$ and $\mu$ 
are all real-valued, the neutralino mass matrix 
\beq
\label{Mneutralino}
M = \left(
  \begin{array}{cccc} M_1 & 0 & -m_Z s_W \cos\beta & m_Z s_W \sin\beta  \\
  0 & M_2 &  m_Z c_W \cos\beta & -m_Z c_W \sin\beta  \\
 -m_Z s_W \cos\beta & m_Z c_W \cos\beta & 0 & -\mu \\
m_Z s_W \sin\beta & -m_Z c_W \sin\beta & -\mu & 0 
\end{array} \right)
\eeq 
is symmetric and can be diagonalized through a 
similarity transformation, i.e.

\begin{equation}
P M P^{-1} = M_{diagonal}
\end{equation}
Analytical results for this diagonalization are well-known \cite{egyptiens}.
However, a straightforward inversion of the expressions of the mass eigenvalues
using these results
is far from obvious. A more manageable alternative is to start from the 
four quantities 
\begin{eqnarray}
&&Tr M \\
&& \frac{(Tr M)^2}{2} - \frac{Tr (M^2)}{2} \\
&& \frac{(Tr M)^3}{6} - \frac{Tr M  \,\,\, Tr (M^2)}{2} + \frac{Tr (M^3)}{3} \\
&& Det M \\
\nonumber
\end{eqnarray}

which are invariant under any similarity transformation. These quantities
allow one to relate the four mass eigenvalues to the initial parameters in 
(\ref{Mneutralino}) 
as follows:

\begin{eqnarray}
&&\tilde{M}_{N_1} + \tilde{M_{N_2} } 
 + \tilde{M}_{N_3} + \tilde{M}_{N_4} = M_1 + M_2   \label{inv1}\\
\nonumber
\end{eqnarray}
\vspace{-1.1cm}
\begin{eqnarray}
&&\tilde{M}_{N_1}  \tilde{M_{N_2} } +
\tilde{M}_{N_2}  \tilde{M_{N_3} }
+ \tilde{M}_{N_3}  \tilde{M_{N_4} }
+ \tilde{M}_{N_4}  \tilde{M_{N_1} }
 =   M_1 M_2 -\mu^2 - M_Z^2 \label{inv2} \\
\nonumber
\end{eqnarray}
\vspace{-1.1cm}
\begin{eqnarray}
&&\tilde{M}_{N_1}  \tilde{M}_{N_2} \tilde{M}_{N_3} 
+ \tilde{M}_{N_2}  \tilde{M}_{N_3} \tilde{M}_{N_4}
+ \tilde{M}_{N_3}  \tilde{M}_{N_4} \tilde{M}_{N_1}
+ \tilde{M}_{N_4}  \tilde{M}_{N_1} \tilde{M}_{N_2} \nonumber \\
&&=   \mu M_Z^2 \sin 2 \beta -(\mu^2 + c_w^2 M_Z^2) M_1 -(\mu^2 + s_w^2 M_Z^2 
        ) M_2 \label{inv3} \\
\nonumber
\end{eqnarray}
\vspace{-1.1cm}
\begin{eqnarray}
&&\tilde{M}_{N_1}  \tilde{M}_{N_2} \tilde{M}_{N_3} 
\tilde{M}_{N_4} = 
 \mu M_Z^2 (c_w^2 M_1 + s_w^2 M_2) \sin 2\beta -\mu^2 M_1 M_2 
 \label{inv4}\\
\nonumber
\end{eqnarray}

Here $\tilde{M}_{N_i}$ denote the eigenvalues of the mass matrix
Eq.(\ref{Mneutralino}), to be distinguished from the physical neutralino
masses ${M}_{N_i}$ given by 
$\tilde{M}_{N_i} \equiv \epsilon_i {M}_{N_i}$, where
$\epsilon_i= \pm$ is undetermined at this level.
 
We emphasize that Eqs.(\ref{inv1}--\ref{inv4}) give
the complete information on the relationship between the four 
neutralino mass eigenvalues and the original parameters,
$\mu$, $M_1$, $M_2$ and $\tan\beta$. Furthermore these equations are a good 
starting point for our purpose, 
%~\footnote{Of course, more complex expressions immediately
%occur once explicitly solving the system of equation (\ref{invariants})
%to extract $M_1$ and the three other neutralino masses, in terms of
%one given physical mass. However, the most complicated equation is of third
%order only.}.  
as they do not favour {\sl a priori} any particular set of parameters.
Thus, the system may be solved in many
different ways, depending on the choice of input/output one is interested in.

According to the algorithm $S_1$, described in the previous section,
 we can assume 
that $\mu$, $M_2$ (and $\tan\beta$) are determined at this stage, 
and extract $M_1$ and the three physical
masses $M_{N_1}$, $M_{N_3}$, $M_{N_4}$ as functions of one mass, say
$M_{N_2}$~\footnote{Actually what we call $M_{N_2}$ plays the role of 
any neutralino mass to be given as input.
There will be a relabeling of neutralino states depending on the values of
other parameters, and one can check afterwards whether the input mass was
the lightest, heaviest or intermediate one.}

Accordingly, in this $S_1$ scenario, one finds for $M_1$

\begin{equation}
 M_1= 
   -\frac{P_{2 i}^2 + 
       P_{2 i} (\mu^2 + M_Z^2 + M_2 S_{2 i} - S_{2 i}^2) 
              + \mu M_Z^2 M_2 s_w^2 \sin 2 \beta}
      {P_{2 i} (S_{2 i} -M_2) + \mu ( c_w^2 M_Z^2 \sin 2 \beta -\mu M_2 )}
\label{solM1}
\end{equation}

where 
$$ S_{2 i} \equiv \tilde{M}_{N_2} + \tilde{M}_{N_i} $$
   
$$ P_{2 i} \equiv \tilde{M}_{N_2}  \tilde{M}_{N_i} $$
and $\tilde{M}_{N_i}$ is any of the three remaining neutralino mass parameters. 

The latter take the following form,

\begin{equation}
\tilde{M}_{N_1}= -\frac{1}{3 a_1} (  a_2 - 2 \sqrt{-A} \cos (Arg[B]))
\label{solMN1}
\end{equation}

\begin{equation}
\tilde{M}_{N_3}= -\frac{1}{3 a_1} ( a_2 + 2 \sqrt{-A} \cos (Arg[B] 
- \frac{\pi}{3} ) )
\label{solMN2}
\end{equation}

\begin{equation}
\tilde{M}_{N_4}= -\frac{1}{3 a_1} ( a_2 + 2 \sqrt{-A} \cos (Arg[B] 
+ \frac{\pi}{3} ) )
\label{solMN3}
\end{equation}

where A, B, $a_1$ and $a_2$ are given in Appendix \ref{app2},
Eqs.(\ref{ais}, \ref{eqB}, \ref{eqA}),
and $Arg[B]$ denotes the phase of the generically complex valued quantity $B$.
Note that $M_1, \tilde{M}_{N_1},\tilde{M}_{N_3}$ and $\tilde{M}_{N_4}$ 
are all real-valued since we chose $\mu$ and $M_2$ to be real.
Note also that Eq.(\ref{solM1}) should yield the same value when 
$\tilde{M}_{N_i}$ is substituted  by any of the three neutralino
masses given by Eqs.(\ref{solMN1} - \ref{solMN3}). This allows a non-trivial
consistency check. For a more detailed discussion of
the derivation of Eqs.(\ref{solM1} - \ref{solMN3}) and related material
the reader is referred to Appendix \ref{app2}. 

\subsection{Combined algorithm scenario $S_2$: 
$M_{\chi^+_1}$, $M_{N_2}$, $M_{N_3}$
input}
Alternatively, when considering now scenario $S_2$
as defined previously
(i.e. the inputs are $\mu$, $M_{N_2}$, $M_{N_3}$),
an even simpler solution occurs for $M_2$ and $M_1$ from 
eqs~(\ref{inv1})--(\ref{inv4}). In fact, the 
necessary and sufficient conditions for the existence of solutions to 
eq.~(\ref{inv1})--(\ref{inv4}) give the  
consistency conditions 
eqs~(\ref{condition1}), (\ref{condition2}) (see Appendix \ref{app2}), 
from which one immediately
obtains, for given $\mu$, simple quadratic 
equations for $M_2$, $M_1$, that we omit to write here. 
The price to pay, as already mentioned previously, is that
without further model--dependent assumption, this combined algorithm for
scenario $S_2$ has to rely on a numerical (iterative) consistency
check, and some non-trivial procedure has to be performed in order not to
miss all possible solutions~\footnote{Indeed, 
solving e.g. eq.~(\ref{invmu2}) for $\mu$ as a function
of $M_{\chi^+_1}$, $M_2$, and bravely injecting the solution
in eqs.~(\ref{condition1})--(\ref{condition2}) gives a highly
non-linear equation for, e.g., $M_2$. Most of the solutions are in fact 
spurious, and our iteration algorithm, based on 
two equations which are only quadratic, copes with these
redundancies in a simpler way.}. 
The upshot is that up to at most {\em four} distinct 
solutions for ($\mu$, $M_1$, $M_2$) 
can occur for given $M_{\chi^+_1}$, $M_{N_2}$,
$M_{N_3}$ input, once all consistency constraints
(including our sign convention choice $M_2 > 0$) are taken into account.
Unfortunately, in contrast with the twofold ambiguities
of algorithm $S_1$ alone, it is not easy to work out
analytically specific domains of the input parameter space 
corresponding to a definite number of distinct solutions, and one
has to rely on the numerics.\\

Let us summarize this section:\\
-in scenario $S_1$, we start from $M_{\chi^+_1}, M_{\chi^+_2} ,M_{N_2}$
to determine first $\mu$ and $M_2$ (up to the possible twofold ambiguity),
and then $M_1$ from 
eq.~(\ref{solM1}), given $\tan\beta$. 
The solution for $M_1$, (\ref{solM1}), is indeed unique for fixed $M_2$, $\mu$ 
and one neutralino mass values.
This part of the algorithm may thus be used in a more general context (than
$S_1$), where $\mu$, $M_2$ and $\tan \beta$ would have been extracted in a way
or the other.
Within scenario $S_1$ alone, there are in principle 
two possible $M_1$ values for any two chargino and one neutralino mass 
input, since the chargino mass input 
does not distinguish the
gaugino-like from the higgsino-like situation, and (\ref{solM1}) is 
not symmetrical under the 
interchange $\vert\mu\vert \leftrightarrow M_2$.\\   
-Alternatively, in scenario $S_2$ one can obtain 
$\mu$, $M_2$, $M_1$ consistently by iteration from  $M_{\chi^+_1}$,
$M_{N_2}$, and $M_{N_3}$.
In practice, this 
combined algorithm converges very rapidly, but a relatively large number
of distinct solutions (up to four) can a priori occur in some
domains of the relevant input parameter space ($\mu$, $M_1$, $M_2$).
[Note that these ambiguities are different from the previous
Higgsino-like $\leftrightarrow$ gaugino-like ambiguity of scenario
$S_1$, the latter  
being precisely removed from the knowledge of a second
neutralino mass]. 

\section{Numerical illustrations of the ino inversion} 
\setcounter{equation}{0}

We shall illustrate here with some representative plots the results
of the inversion in the ino sector, according to the algorithms
explained in section 3, together
with a few  remarks and comments. Since our choice
of input masses is rather arbitrary and may not directly 
correspond to the most interesting experimental situation, our
comments are accordingly essentially qualitative. We will see,
nevertheless, that a number of general and interesting properties
of the inversion can directly be seen here, irrespective of the
precise values of the other parameters that have to be fixed,
like $\tan\beta$ typically.

\subsection{Scenario $S_1$: two charginos plus one neutralino input}
 
\begin{figure}[htb]
\epsfxsize=120mm
\epsfysize=120mm
\begin{center}
\hspace*{0in}
\epsffile{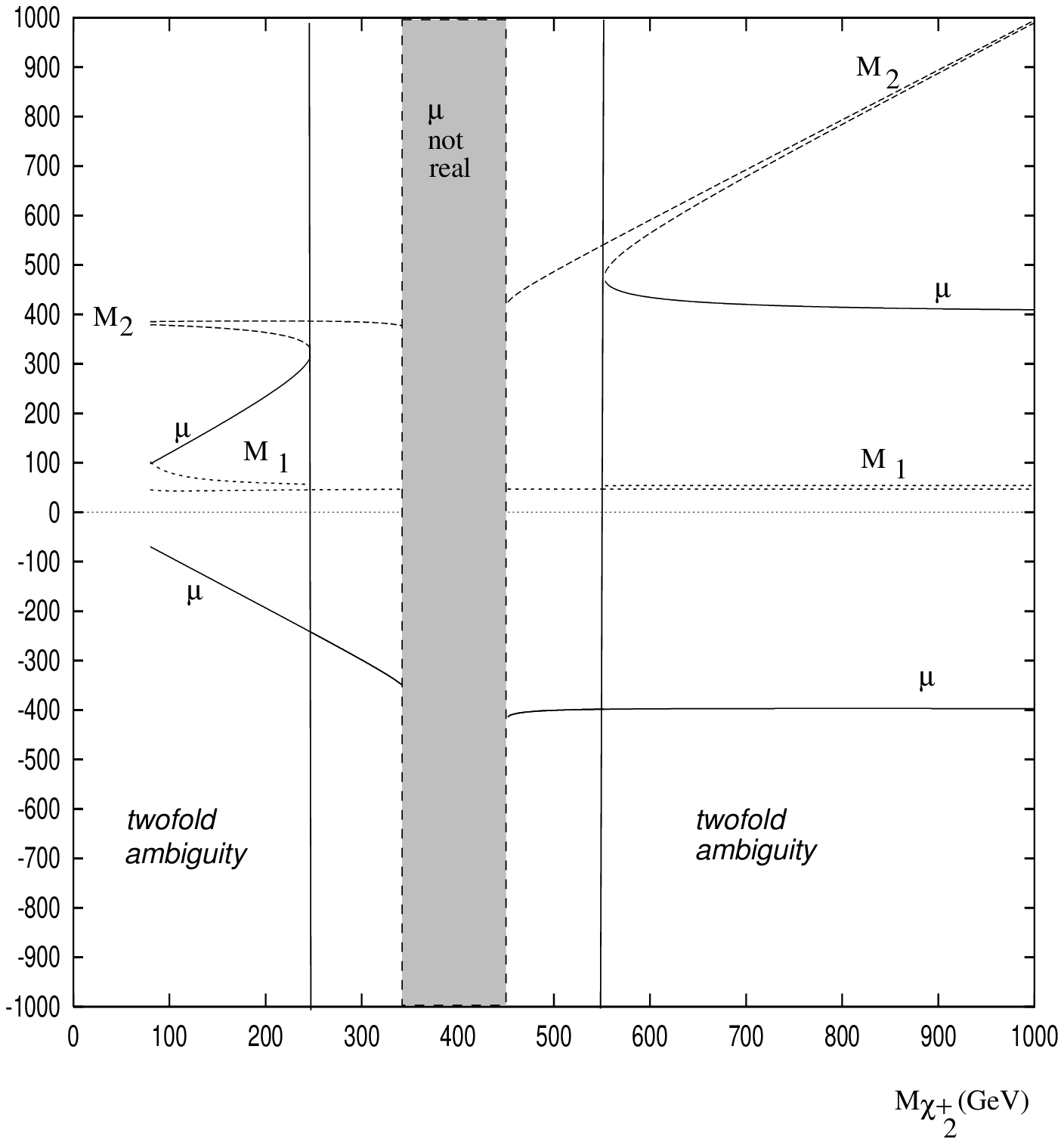}
\vspace*{0.1in}
\caption{ \label{fig2} $\mu$, $M_2$ and $M_1$ (with the
``higgsino-like" convention
$\vert\mu\vert \leq M_2$) as functions of  
$M_{\chi^+_2}$ for fixed $M_{\chi^+_1}$ ( = 400 GeV);  
$M_{N_2}$ ( = 50 GeV), and $\tan\beta$ ( = 2). The opposite
``gaugino-like" situation, with $\vert\mu\vert \geq M_2$, is trivially
obtained by the substitution $M_2 \to \vert \mu \vert$ and
$\mu \to \mbox{sign}(\mu) M_2$.}
\vspace*{-0.1in}
\end{center}
\end{figure}

We first illustrate the basic algorithm $S_1$, namely
the reconstruction of 
$\mu$, $M_2$ (and $M_1$) from $M_{\chi^+_1}$, $M_{\chi^+_2}$ (and
$\tan\beta$ and $M_{N_2}$) in fig. \ref{fig2}. 
As already mentioned,   
we illustrate only the higgsino-like situation
$\vert\mu\vert \leq M_2$ since the gaugino-like illustration is trivially
obtained from it (see captions of fig. \ref{fig2}). 
 
Since it is phenomenologically more likely~\cite{SNOWMASS}
that the first inos discovered (if any) 
will be two neutralinos and
only one chargino (the second chargino 
being of heavier mass),
we fixed in scenario $S_1$ only one chargino mass,
say $M_{\chi^+_1}$, while varying the other one, $M_{\chi^+_2}$, in
a large range
to illustrate as much as possible the
dependence on the physical input   
 (see figure captions). The shape of the various
plots in fig. 2 is quite generic. First, it exhibits three
distinct zones as regards the existence, uniqueness, or possible
ambiguities on $\mu$, $M_2$, $M_1$ (see also the discussion and table 1 in
section 3): 
\begin{itemize}
\item[ i)] the grey shaded region, where there are no solutions for
real $\mu$, corresponds to zone (I) of table 1, with the corresponding 
constraint on $\sin 2 \beta$ not fulfilled~\footnote{Of course, more generally one 
could be interested in complex $\mu$ solution. The results of
section 3.2 and appendix B, not directly applicable in this case,
would then need a generalization which is beyond the scope of the present 
paper.}. 
If one takes a smaller or larger $M_{\chi^+_1}$, this
region around $M_{\chi^+_1}$ will be simply displaced accordingly. 

\item[ ii)] in the left and right 
boarder zones are the regions of twofold ambiguities
on $\mu$, $M_2$, as indicated. Note therefore that one of the two solutions
has a discontinuity at the boarder between the single and twofold solutions
region. Without additional information (or particular model assumption) one 
cannot a priori reject any of the two solutions.

\item[ iii)] Finally the two bands in between zones i) and ii) correspond to
the region  where
eqs. (\ref{invmu2}, \ref{invM2}) give a unique solution for $\mu$ and 
$M_2$. It is interesting to note that those bands are narrower
when $\tan\beta$ is increasing (in fig. 2 
$\tan\beta =2$), irrespective of the $M_{\chi^+_1}$ values, 
becoming e.g. only a few GeV wide for $\tan\beta > 35$. 
\end{itemize}
Moreover, as a general behaviour, the 
values of $\mu$ and $M_2$ are rather insensitive
to $\tan\beta$ (apart obviously from the discontinuous change 
occurring for one of the solution at the boarder between zones ii) and iii)).
One can also remark from fig. 2
 the relatively simple shape of $\mu$ and $M_2$ as function of 
 $M_{\chi^+_2}$, 
with an almost constant or linear dependence, apart from some narrow
regions. This is easily understood, since from eqs.~(\ref{invmu2}), 
(\ref{invM2}) one obtains $(\mu , M_2) \simeq ( M_{\chi^+_2}, M_{\chi^+_1})$,
(resp. $ (M_{\chi^+_1}, M_{\chi^+_2}) $ ) for $M_{\chi^+_2} << M_{\chi^+_1}$
(resp.  $M_{\chi^+_2} >> M_{\chi^+_1}$).\\

In fig. \ref{fig2} 
we also plot $M_1$ for the corresponding values of $\mu$ and
$M_2$, and for fixed $M_{N_2}$. The twofold valuedness of $M_1$ there is
entirely due to the twofold ambiguity of $\mu$ and $M_2$. In contrast,
we do not illustrate here the other determination of $M_1$ resulting
from the
$\vert\mu\vert \leftrightarrow M_2$ interchange: although the resulting
$M_1$ may be in general quite different, this interchange will have 
very little effects as a function of $M_{\chi^+_2}$. First of all, 
the trivial, almost constant behaviour of $M_1$ in this
plot (except for small $M_{\chi^+_2} \lsim $ 100 GeV)
is relatively simple to understand from eq.~(\ref{solM1}). 
%Since
%($M_2$, $\vert\mu\vert$) $\simeq$ ($M_{\chi^+_1}$, $M_{\chi^+_2}$) for 
%large enough $M_{\chi^+_i}$, 
Indeed, $M_2$ is large for large $M_{\chi^+_2}$ (since we assume $\vert \mu \vert < M_2$ ). In this limit
and for fixed $M_{\chi^+_1}$, $M_{N_2}$ (and $\tan\beta$) one finds from
eqs.(\ref{ais}-\ref{sol3}) that the remaining three neutralinos behave
like $M_2$. One then easily determines the (constant) behaviour of $M_1$ from 
eq.(\ref{solM1})
in the large $M_2$ limit, namely
$ M_1\simeq  [M_{N_2} (M_{N_2}^2 
- \mu^2 - m_Z^2) - s_w^2 \mu \; m_Z^2 \sin 2 \beta ]/(M_{N_2}^2 - \mu^2)$
%The only relevant variation
%of $M_1$ in eq. (\ref{solM1}) is from $M_2$ , which in
%both cases gives asymptotically $M_1 \simeq const$, for large enough
%$M_2$ (or $\vert\mu\vert$). \\
    
In summary,  
the information from the plots in fig. 2 is rather interesting:
apart from some small regions,
the dependence of $\mu$, $M_2$ (and even $M_1$ to
some extent) upon $\vert M_{\chi^+_1} -
M_{\chi^+_2} \vert $ is very simple
for a wide range of the latter mass difference. In other words,
once one chargino mass will be known with some accuracy 
(together with $\tan\beta$),
fig. 2 indicates that, in principle, the possible values of $\mu$,
$M_2$, and $M_1$ are strongly correlated.
 Moreover, although in fig 2 we have only
varied $M_{\chi^+_2}$ the behaviour is quite generic and does not
change much, qualitatively,  when the other fixed 
inputs ($M_{N_2}$ and $\tan\beta$) are varied. \\

\begin{figure}[htb]
\epsfxsize=120mm
\epsfysize=120mm
\begin{center}
\hspace*{0in}
\epsffile{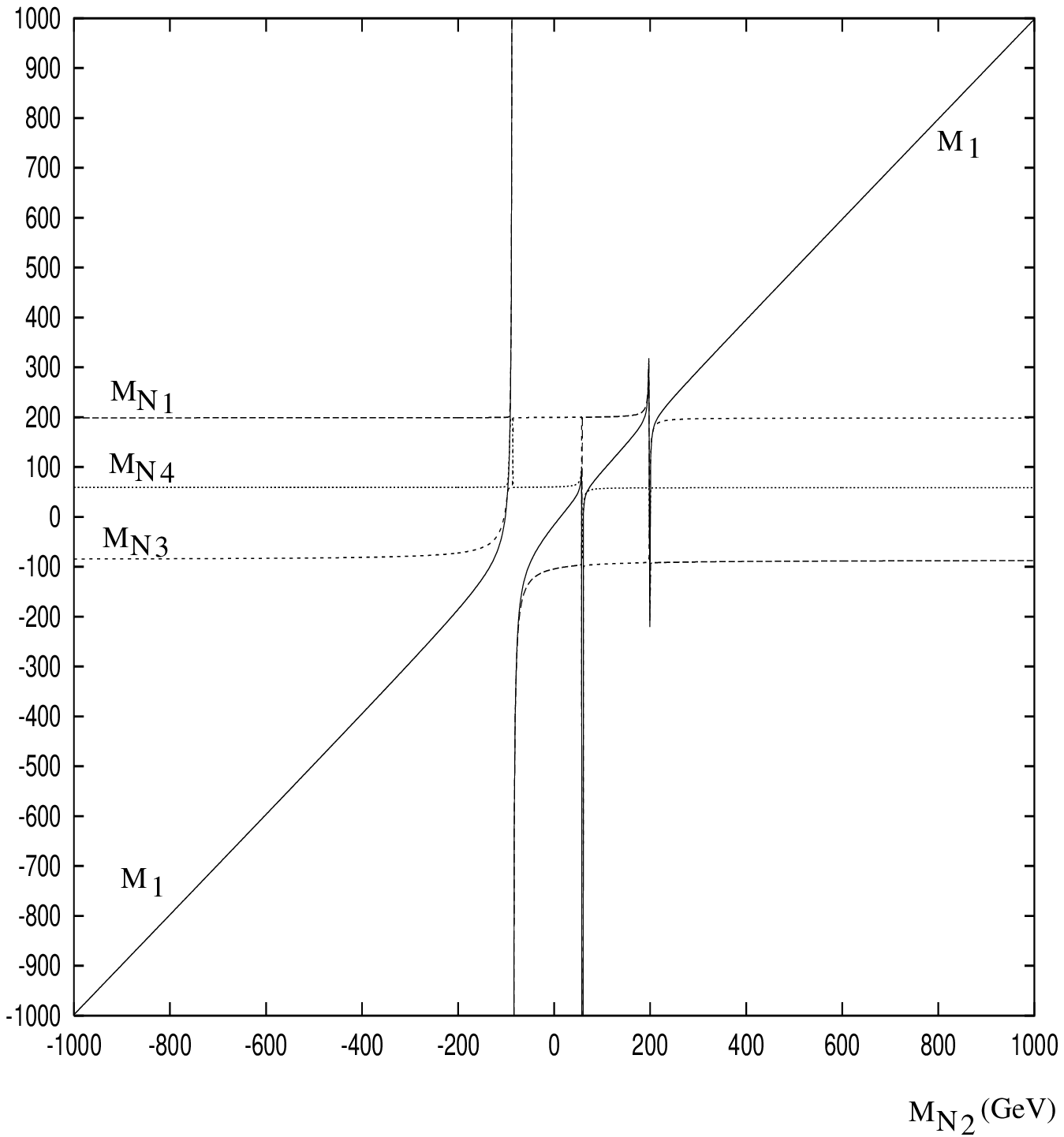}
\vspace*{0.1in}
\caption{ \label{fig3} $M_1$ and the three neutralino masses 
($M_{N_1}$, $M_{N_3}$, $M_{N_4}$) as functions of $M_{N_2}$ for fixed
$M_{\chi^+_1}$ (= 80 GeV), $M_{\chi^+_2}$ (= 200 GeV) and 
$\tan\beta$ (= 2).  }
\vspace*{-0.1in}
\end{center}
\end{figure}

Finally, as a last illustration of scenario $S_1$, we plot 
in fig. \ref{fig3} the resulting values of $M_1$ and the three other 
neutralino masses, $M_{N_1}$, $M_{N_3}$, $M_{N_4}$, determined from
eqs.~(\ref{solMN1})--(\ref{solMN3}). Note that the plots in fig. \ref{fig3}
are now functions of $M_{N_2}$, and with
different $M_{\chi^+_1}$, $M_{\chi^+_2}$ inputs than for fig. \ref{fig2}. 
The singularities of $M_1$ for specific values of $M_{N_2}$,
correspond simply to the denominator of eq.(\ref{solM1}) vanishing
(and also correspond to the exchange among neutralino masses, as
illustrated on the plots). Apart from relatively small regions
around the singularities, the almost linear dependence of 
$M_1$ as a function of $M_{N_2}$ 
can be traced back to the
fact that $M_1$  behaves, for asymptotically large 
$M_{N_2}$ in eq.(\ref{solM1}), as
\begin{equation}
M_1 \sim M_{N_2} + \frac{1}{M_{N_2}} [ M_{N_i} ( M_{N_i} - M_2) - \mu^2 - M_Z^2
+ \frac{\mu}{M_{N_i}} ( \mu M_2 - M_W^2 sin 2 \beta)] + O(\frac{1}{M_{N_2}^2})
\label{asymptotM1}
\end{equation}
where $M_{N_i}$ is any of the neutralino masses other than $M_{N_2}$. 
Strictly speaking, 
 (\ref{asymptotM1}) is valid only if the $M_{N_i}$'s become insensitive
to the value of $M_{N_2}$ in this limit. That this is true can be easily seen
from the fact that all the $a_i$'s in Eq.(\ref{ais}) behave in this limit like
$\sim M_{N_2}^3$. Then the dependence on $M_{N_2}$ tends to cancel out in 
Eq.(\ref{cubiceq}), making the three neutralino masses insensitive to this
parameter, as illustrated in fig. \ref{fig3}. 

\subsection{GUT scale  values of the ino mass parameters}

\begin{figure}[htb]
\epsfxsize=130mm
\epsfysize=130mm
\begin{center}
\hspace*{0in}
\epsffile{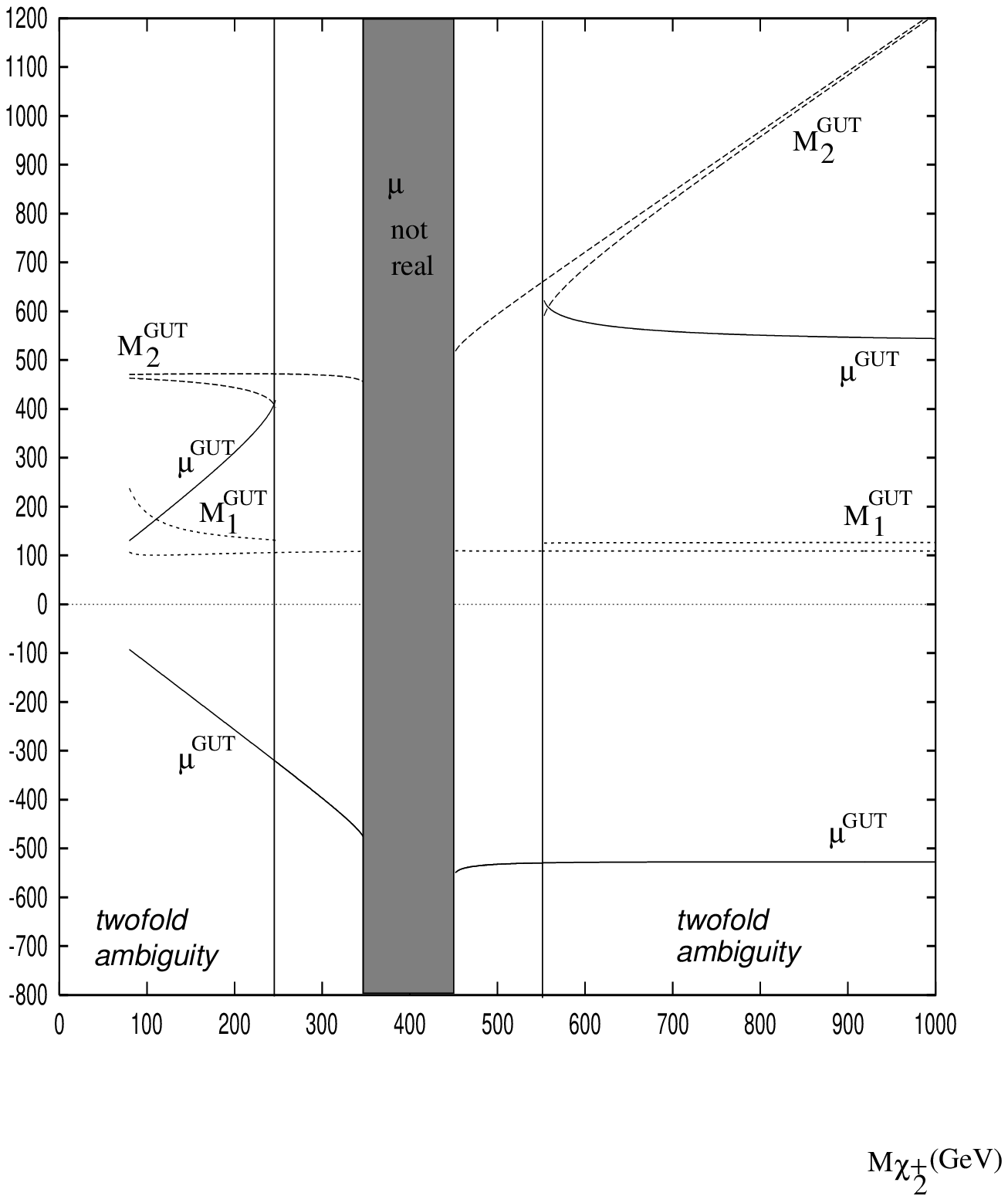}
\vspace*{0.1in}
\caption{ \label{fig4} Same captions as for fig. 2 but with 
$\mu$, $M_2$ and $M_1$ evolved up to GUT scale. }
\vspace*{-0.1in}
\end{center}
\end{figure}

Next, in fig. \ref{fig4} we 
illustrate plots very similar to the ones in fig. \ref{fig2},
but where in addition the parameters $\mu$, $M_2$, $M_1$ have been 
evolved from a low scale, $Q_{low}\sim $ 200 GeV, to a GUT scale, $Q_{GUT}\sim
2 \times 10^{17}$ GeV 
according to the renormalization group
evolution (RGE) \cite{rge}. 
%
% additionnal remarks on RGE starts here:
%
In order to illustrate  as simply as possible
some representative values 
after RG evolution, note that we have made a number of approximations for
the evolution itself, that we feel are reasonable for the present purpose.
Namely, the RG evolution
is limited to the one-loop approximation,
and we have assumed in addition a
single, universal supersymmetric threshold, that
is identified with the low energy scale, $Q_{low}\sim $ 200 GeV,
where the evolution starts (so that the RG evolution is dictated by fully
supersymmetric beta function coefficients between $Q_{low}$ and $Q_{GUT}$). 
There should be no difficulties in principle to
incorporate in our framework more realistic supersymmetric 
threshold effects, provided however that the masses of other partners
(Higgses and sfermions) are known at this stage in addition to 
ino masses.
The latter refinements should however produce relatively small
corrections to the numerical values that are illustrated 
in fig. \ref{fig4}.\\

It should not be
very surprising that RGE scarcely change the shape of the various plots, 
with respect to those in fig. \ref{fig2},
apart from a systematic shift and a slight change in the slopes.
This is a direct consequence of the generic form of the one-loop
evolution equations for the gaugino masses and the $\mu$ parameter,
namely $$\frac{dm}{dt} \sim m \times \sum_{i}{\cal G}_i^2$$
where ${\cal G}_i$ refers to
either gauge or Yukawa couplings and $t$ is the evolution parameter. 
Taking into account
the evolution of the couplings themselves, one then expects the mass
parameters to behave like $$m(t) \sim m_0 \times 
\prod_{i}{\cal G}_i(t)^{a_i}/{\cal G}_i(0)^{a_i}$$ 
where $a_i$ is some numerical power and $m_0$ the mass parameter at some initial 
scale.[The latter behaviour is exact, at the one-loop level,
for the gauge couplings and also
for the Yukawa couplings at least in the small $\tan \beta$ regime.]
Thus the only effect of the running is a coupling dependent rescaling of the 
inversion results.  

%apart from a systematic shift. This is simply due to the relatively
%trivial (one-loop) RG behaviour of the gaugino masses, 
%and of the $\mu$ parameter: the first order beta function coefficients 
%for those parameters are such that they evolve independently of each others.
%In other words,   
%the $M_2$, $M_1$ shift due to
% RGE from fig \ref{fig2} to fig. \ref{fig4} is practically 
%independent of the $\mu$, $M_2$, $M_1$ values.\footnote{An apparent exception
%is the relatively more pronounced evolution of $M_1$ as compared
%to $M_2$ or $\mu$, in particular
% for small $M_{\chi^+_2}$: this is simply due to the relatively
%large value of $\vert b_1 \vert$,
%the one-loop beta coefficient of the gauge coupling $g_1$, which is 
%essentially driving the RGE of $M_1$~\cite{rge}.}. \\

\subsection{Scenario $S_2$: one chargino plus two neutralinos input}
\begin{figure}[htb]
\epsfxsize=120mm
\epsfysize=120mm
\begin{center}
\hspace*{0in}
\epsffile{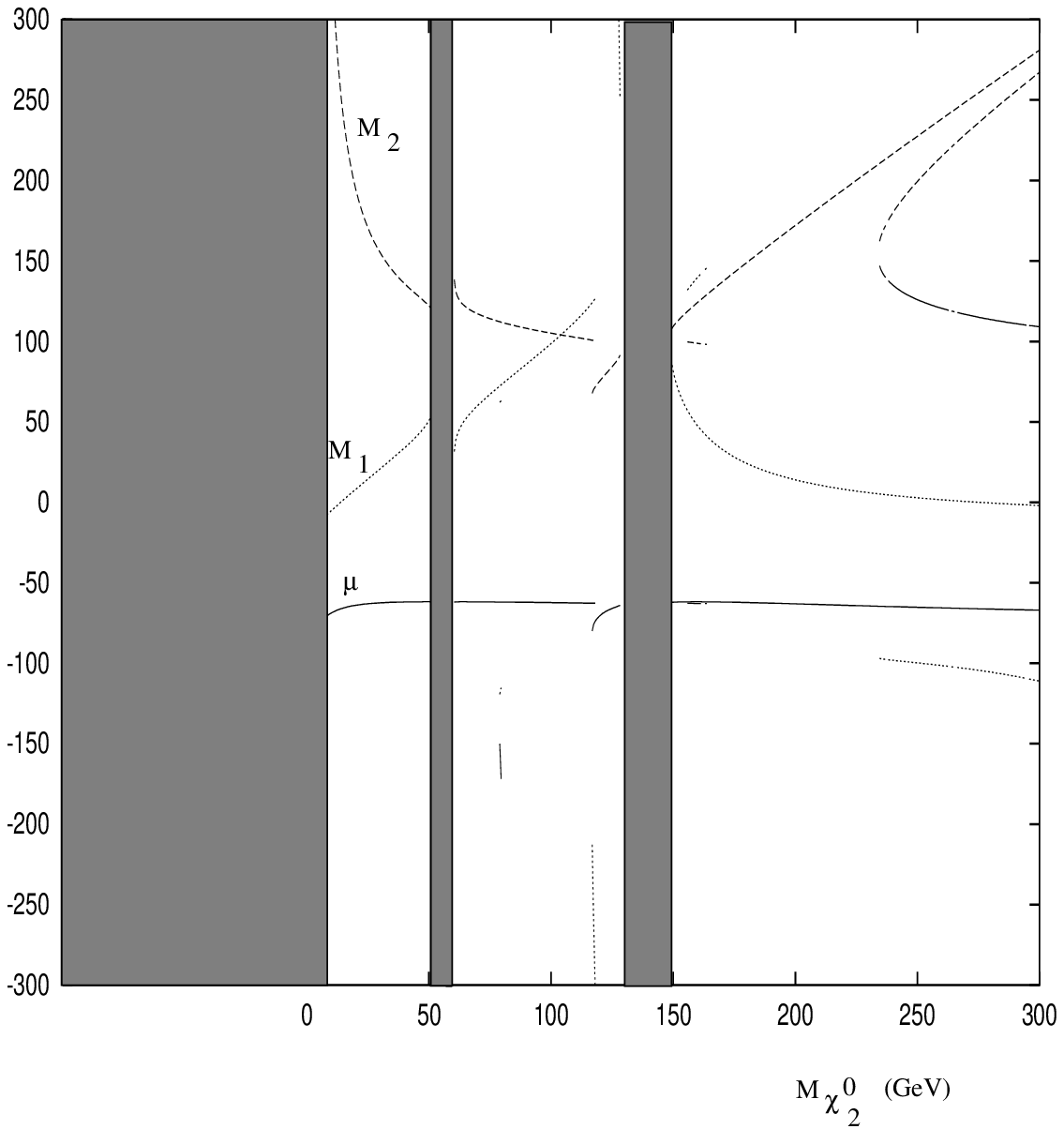}
\vspace*{0.1in}
\caption{ \label{fig5} $\mu$, $M_2$ and $M_1$ as function of  
$M_{\chi_2}^0 \equiv M_{N_2}$ 
for fixed $M_{N_3}$ ( $= -100 $GeV), $M_{\chi^+_1}$ 
(= 80 GeV)  
and $\tan\beta( = 2$).}
\vspace*{-0.1in}
\end{center}
\end{figure}

Next we illustrate the (probably  
phenomenologically more relevant) scenario
$S_2$, namely
with $M_{\chi^+_1}$,  $M_{N_2}$ and $M_{N_3}$ 
as input. As expected, fig.~\ref{fig5} reflects
the more algebraically involved inversion from the  
combined algorithm $S_2$
(with unknown $ M_{\chi^+_2}$), with the possible
occurrence of several distinct solutions for ($\mu$, $M_1$, $M_2$), 
as discussed previously. 
However, apart from the relatively untidy behaviour in some narrow
zones, the domains of unique and twofold (or more) solutions 
are relatively well separated over a wide range of the $M_{N_2}$ values.
Moreover, the behaviour of the plots for the  
particular input values in fig. \ref{fig5} turns out to be quite generic. 
The shaded regions again corresponds to a 
zone where at least one  
of the output parameters ($\mu$, $M_1$, $M_2$) 
becomes complex-valued. 
In fig.~(\ref{fig5}) we only show 
on purpose a range of values such that all masses are relatively
light, while for larger  $M_{N_2}$ the dependence of
$\mu$, $M_2$, $M_1$ upon the latter 
becomes simpler, almost linear,  in 
accordance with the behaviour in the 
previous figures for scenario $S_1$ alone. There are also specific values
of the input masses such that one of the solutions for 
$\mu$, $M_1$, or $M_2$ is becoming very large, due to an explicit pole 
in the analytic inversion, as reflected e.g. 
in one of the $\mu$ and $M_2$ solutions, 
for $M_{N_2}\simeq$ 80 GeV and $M_{N_2}\simeq$ 123 GeV, respectively. \\
Also, the dependence of scenario $S_2$ upon
$\tan\beta$ variations is relatively mild. In contrast,
varying $M_{\chi^+_1}$ and/or $M_{N_3}$ 
input values   
can have more drastic effects, since in particular
the occurrence of multi-fold solutions  
depends on those inputs. 

\begin{figure}[htb]
\epsfxsize=120mm
\epsfysize=120mm
\begin{center}
\hspace*{0in}
\epsffile{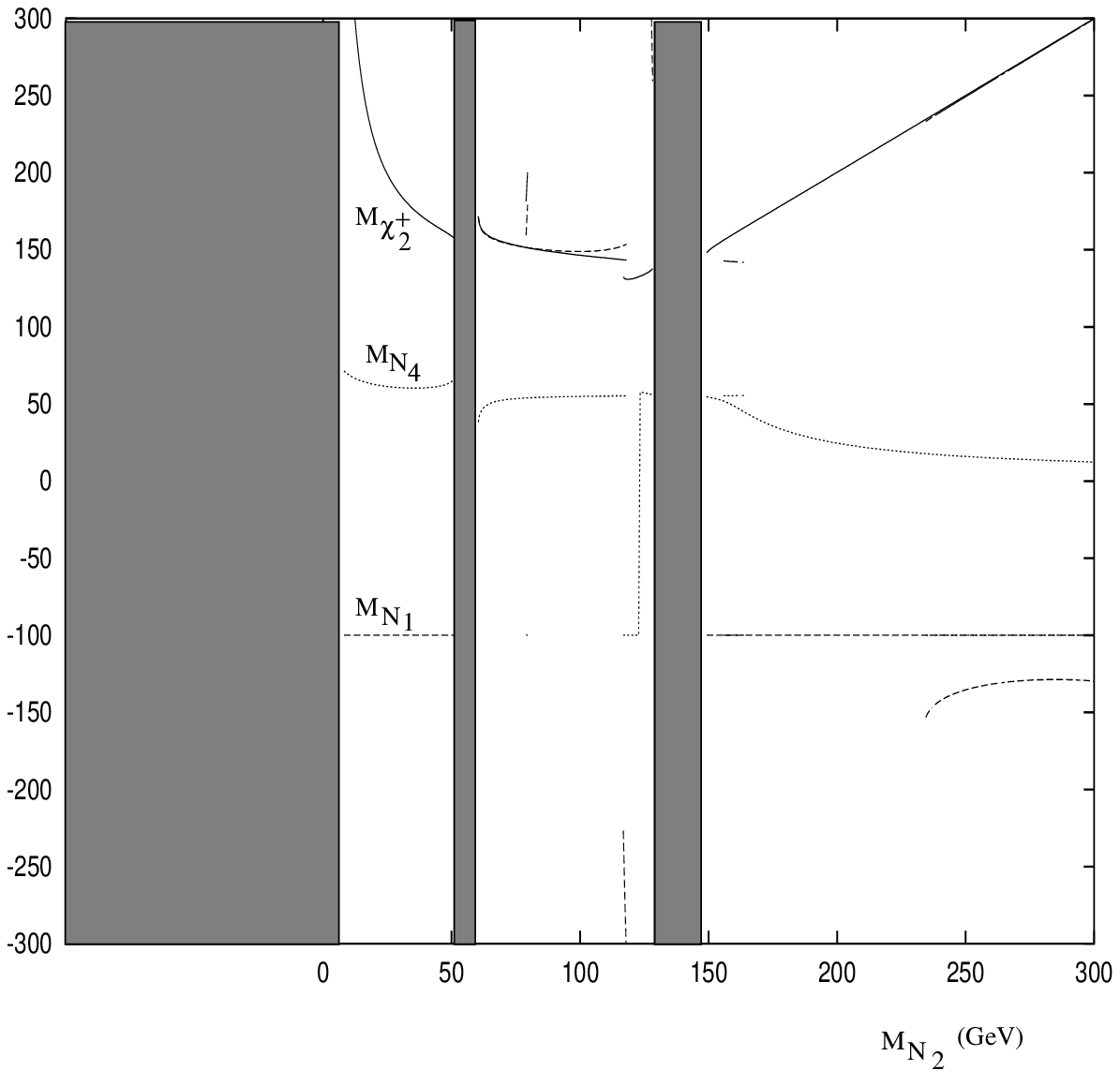}
\vspace*{0.1in}
\caption{ \label{fig6} resulting $M_{\chi^+_2}$, $M_{N_1}$ and $M_{N_4}$ 
values for the 
$\mu$, $M_2$, $M_1$ values of fig.~\ref{fig5}.}
\vspace*{-0.1in}
\end{center}
\end{figure}

Finally, for completeness, we show in fig.~\ref{fig6} the 
values of the
other physical masses $M_{\chi^+_2}$, $M_{N_1}$ and $M_{N_4}$, 
resulting from the same input choice as in fig. \ref{fig5}. 
Note that, apart from $M_{N_1}$, the twofold valuedness of 
$M_{\chi_2}^+$ and $ M_{N_4}$ for $M_{N_2} \gsim $230 GeV
turns out to be numerically negligible for this particular input.
In a more complete
and realistic analysis, it should be possible
 to examine e.g. whether some of the
multi-fold solutions in fig.~\ref{fig5} could be excluded by looking
at the consistency of the resulting other physical ino masses
with data. 

\section{Other soft breaking parameters inversion}
\setcounter{equation}{0}
%Once (or before) the gaugino sector soft-breaking parameter space
%has been reconstructed from the physical masses, it is natural
%to attempt such a reconstruction for the remaining part of the soft-breaking
%Lagrangian. 

Ultimately, one would need to reconstruct the remaining part of the 
soft-breaking Lagrangian, i.e. the sfermion and Higgs sectors.
In contrast with the ino sector, 
the de-diagonalization of these sectors 
would not present much analytical difficulties (at least at the tree-level),
provided one knows a sufficient number of physical masses. 
%A possible physical mass measurement strategy for the sfermion and/or
%Higgs sector of the MSSM is however beyond the scope of the present paper.  
%Accordingly we simply sketch here for definiteness some
A full strategy (including loop effects) is, however, beyond the scope of
the present paper.
Our aim here is to simply sketch for completeness some
straightforward inversion formulas and make some comments.

\subsection{Sfermion parameter inversion}
In the sfermion
sector, de-diagonalization is straightforward
since it involves $2 \times 2$ mass matrices if neglecting flavor
non-diagonal terms. For instance,
from
eqs.~(\ref{Mstop}), and assuming the two physical stop 
masses $m_{\tt_1}$, $m_{\tt_2}$ (with 
the convention $m_{\tt_1} <  m_{\tt_2}$) and the mixing angle 
$\theta_{\tt}$ to be given as input (plus $\tan\beta$), 
one immediately obtains
\beq
A_t =\frac{\mu}{\tan\beta} +(m^2_{\tt_2}-m^2_{\tt_1}) 
\frac{\sin 2\theta_{\tt}}{2\;m_t}
\eeq 
and
\beqn
M^2_Q =  m^2_{\tt_1} \cos^2\theta_{\tt}+  
m^2_{\tt_2} \sin^2\theta_{\tt} -m^2_t -
\cos(2\beta)\;(4 m^2_W-m^2_Z)/6 \\ \nn
M^2_R =  m^2_{\tt_1} \sin^2\theta_{\tt}+  
m^2_{\tt_2} \cos^2\theta_{\tt} -m^2_t +\frac{2}{3}
\cos(2\beta)\;(m^2_W-m^2_Z)
\eeqn
Similar expressions are obtained for the sbottom and stau parameters.\\
Evidently, in the case of a more constrained scenario, one would not
need all sfermion physical masses and mixing angles to extract the relevant
soft terms. In fact,  
even in the unconstrained MSSM case, $M_Q$ is a common parameter to
stop and sbottom mass matrices , so that 
to reconstruct the complete soft terms of this 
sector one ``only" needs five parameters, to be chosen 
 among the four masses and two mixing angles. Conversely, notice that
if knowing
the six previous physical input, one can
determine $tan\beta$ very simply  from 
\beq
\label{tbetasf}
m_W^2 \cos 2\beta = m^2_{\tt_1} \cos^2\theta_{\tt}+  
m^2_{\tt_2} \sin^2\theta_{\tt} -m^2_{\tb_1} \cos^2\theta_{\tb} 
-m^2_{\tb_2} \sin^2\theta_{\tb} +m^2_b -m^2_t \;.
\eeq

\subsection{Higgs sector inversion}
Going from physical masses to soft SUSY--breaking
parameters in the Higgs sector also does not involve much 
difficulties, at least naively. In fact,   
electroweak symmetry breaking (EWSB) gives two 
(necessary but not sufficient) constraints,
commonly used  
to fix the $B$ and $\mu$ parameter at the EWSB 
scale:
\beqn
\label{EWSB}
\mu^2 = \frac{m^2_{H_d} -\tan^2\beta \; m^2_{H_u}}{\tan^2\beta -1} 
-\frac{m^2_Z}{2}\;; \\
\nn
B \mu = (m^2_{H_d} +m^2_{H_u} +2\mu^2)\sin\beta \cos\beta\;.
\eeqn 

Now, one may assume for example that $\mu$
is extracted from the previous ino inversion, and
use eqs. (\ref{EWSB}), together with the definition of the physical
Higgs masses, eqs.~(\ref{Hmass}), to determine
the basic Higgs parameters.
Typically, given the lightest CP-even higgs mass $M_h$,
one obtains for the soft parameters at the tree-level:
\beq
\label{MHuHd}
 m^2_{H_u} =\frac{M^2_A-(\mu^2+m^2_Z/2)(\tan^2\beta-1)}
{1+\tan^2\beta} \;\;\;;\;\; 
m^2_{H_d} = M_A^2-M^2_{H_u} -2 \mu^2
\eeq

\beq
\label{Binv}
B = (m^2_{H_d}+m^2_{H_u} +2\mu^2) \frac{\sin(2\beta)}{2\mu}
\eeq
and
\beq
\label{mAeq}
M_A^2= \frac{M_h^2 ( m_Z^2 - M_h^2)}{m_Z^2 \cos^2 2 \beta - M_h^2}
\eeq 
and the other physical Higgs masses are determined from eqs.~(\ref{Hmass}).
Of course as is well known radiative corrections 
to the Higgs masses (in particular that of the lightest CP-even)
are very important to take into account~\cite{HiggsRC,CaWaQui}, 
for some values of the sfermion masses and mixing angles, so that it is 
much too unrealistic to restrict to such a tree-level Higgs parameter
inversion. Including the leading one-loop corrections
in expressions (\ref{MHuHd} - \ref{mAeq}) is in fact 
manageable.
%More precisely, one  
%may calculate the contributions to the one-loop effective potential
%in (\ref{VHiggs}),
%obtaining  corrections 
%$\Delta V_{d,u} \equiv \partial V^{1-loop}/\partial v_{d,u}$ to the
In this case the induced dependence on the full parameter set of
the MSSM should be in principle taken into account. In particular, the
parameters of the Higgs sector we are solving for enter now in a more
complicated way equations (\ref{EWSB}). However, a very good starting point
would be to consider only the dominant top/stop-bottom/sbottom contributions   
in which case the corrections to Eqs.(\ref{MHuHd} - \ref{mAeq}) can still
be written in an analytical form. Otherwise, one will still 
have to resort to a thoroughly numerical procedure.

\section{Discussion and outlook}
In this paper we have worked out a systematic 
inversion algorithm and strategy
to obtain from relatively simple algebraic relations 
the basic MSSM Lagrangian, 
from some appropriate physical mass inputs.
We mostly concentrated on the ino parameter inversion, which is
the less straightforward algebraically,
 due to a complicated diagonalization structure.
The advantage over possibly more direct methods 
(like e.g. some systematic scanning of the Lagrangian parameter space),
is that it gives a fast, reasonably tractable algorithm, and also
clearly points out to the existence 
of non-trivial ambiguities in such a reconstruction, occurring
in some definite range of the input mass values~\footnote{Our main algorithm 
will be soon available as a computer fortran code, on request to the 
authors.}.\\
In addition, our results exhibit the relatively simple and
generic behaviour of
most of the Lagrangian parameters as functions of input masses,
apart from some narrow regions where the behaviour is more involved.
If developed further, performing more systematic simulations,
this approach may thus 
potentially give useful insight into the precise connection between  
the basic MSSM parameters and the experimentally measured ones.   
In the most optimistic case (that is, if knowing a sufficient 
number of  physical input ino masses
and with sufficient accuracy)
it allows for a precise reconstruction of the
unconstrained MSSM ino sector Lagrangian. In addition, some of the 
theoretically well-motivated additional assumptions, like for instance the
universality of soft mass parameters, should be easily implementable
in our basic algorithm, due to its relative flexibility. 
%Specific
%SUSY model--building assumptions are beyond the scope of the present
%paper, but more concretely we shall simply mention
%that, for instance, the well-motivated 
Typically, gaugino mass parameter universality
at the GUT scale, leading to the approximate relation
$M_1(M_Z) \simeq 5/3 \tan^2\theta_W M_2(M_Z)$, can be 
very simply combined with our iterative algorithm $S_2$, to 
give now $\mu$, $M_2$ and $M_1$ 
in terms of only one chargino
and one neutralino mass.\\

Finally we also mention the complementarity of our approach with 
the one discussed recently in ref.~\cite{choi}. In particular, our simple
analytical determination of $M_1$, eq. (\ref{solM1}), may be readily 
implemented as well in their construction. Both procedures 
are however purely theoretical, 
in the sense that the influence of experimental 
errors on the input parameter measurements have not yet been 
taken into account. In any case, clearly an algebraic approach 
is a useful tool for a more systematic phenomenological study
(taking into account e.g. the present bounds or the 
expected future accuracy on input parameters) to be presented  
elsewhere~\cite{prepar}.

\noindent {\bf Acknowledgments:} \\
We thank Abdelhak Djouadi 
for useful discussions. 
This work is partially supported by the French GDR--Supersym\'etrie.

\appendix
\section {MSSM mass formulas}\label{app1}
\setcounter{equation}{0}
We collect here for completeness different useful expressions for the 
sfermion,
ino and Higgs mass matrices or eigenmasses.

--Gaugino sector: \\
the chargino mass matrix reads from Eqs.~(\ref{Lgaugino}, \ref{Lneutralino})
\beq
\label{Mchargino}
{\cal M}_C = \left(
  \begin{array}{cc} M_2  & \sqrt 2 m_W \sin\beta  \\
              \sqrt 2 m_W \cos\beta  & \mu  
\end{array} \right)
\eeq 
with the chargino mass eigenvalues:
\beq
\label{Mchi12}
M^2_{\chi_{1,2}} = \frac{1}{2} [ M^2_2 +\mu^2 +2 m^2_W \pm 
[ (M^2_2-\mu^2)^2 +4 m^4_W \cos^2 2\beta +4 m^2_W (M^2_2 +\mu^2
+2 M_2\mu \sin 2\beta)]^{1/2} ] \;;
\eeq
and the diagonalization of the non-symmetric matrix (\ref{Mchargino}) involves
two mixing angles, $\phi_\pm$:
\beq
\label{phi-}
\tan 2\phi_- = 2\sqrt 2 m_W \frac{\mu \sin\beta +M_2 \cos\beta}
{M^2_2-\mu^2 -2m^2_W \cos 2\beta}
\eeq
\beq
\label{phi+}
\tan 2\phi_+ = 2\sqrt 2 m_W \frac{\mu \cos\beta +M_2 \sin\beta}
{M^2_2-\mu^2 +2m^2_W \cos 2\beta}
\eeq

--Higgs sector: \\
the mass eigenvalues of the neutral CP-odd A, CP-even $h$, $H$  and 
charged $H^+$ scalars read:
\beqn
\label{Hmass}
M^2_A & =&m^2_{H_u}+m^2_{H_d} +2\mu^2 +\Delta_A \;; \\ \nn
M^2_{H\pm} & =& M^2_A +m^2_W  +\Delta_{H+} \;; \\ \nn
M^2_{h,H} &  =&\frac{1}{2} [ M^2_A +m^2_Z \mp [ (M^2_A+m^2_Z)^2 -4 M^2_A m^2_Z
\cos^2 (2\beta) ]^{1/2} ] +\Delta_{h,H} \;; 
\eeqn
where $\Delta_A$, $\Delta_{H+}$, and $\Delta_{h,H}$ denote radiative
correction contributions, whose complicated expressions in one-loop
approximations are given e.g. in refs.~\cite{CaWaQui, HmassRC}. \\

--Sfermion sector: \\
The mass matrices read 
\beq
\label{Mstop}
{\cal M}^2_{\tt} = \left(
  \begin{array}{cc} M^2_Q + m_t^2+(\frac{2}{3}m^2_W-\frac{1}{6}m^2_Z)
\cos 2\beta  & m_t \, (A_t -\mu/\tan\beta) \\
               m_t (A_t -\mu/\tan\beta) & m_{t_R}^2 + m_t^2 - 
\frac{2}{3}(m^2_W-m^2_Z)\cos 2\beta
\end{array} \right)
\eeq 
\vspace{1cm}
\beq
\label{Msbottom}
{\cal M}^2_{\tb} = \left(
  \begin{array}{cc} M^2_Q + m_b^2 -(\frac{1}{3}m^2_W+\frac{1}{6}m^2_Z)
\cos 2\beta   & m_b \, (A_b -\mu \tan\beta) \\
               m_b (A_b -\mu \tan\beta) & m_{b_R}^2 + m_b^2 + 
\frac{1}{3}(m^2_W-m^2_Z)\cos 2\beta
\end{array} \right)
\eeq 
which after diagonalization give the stop and sbottom mass eigenvalues
$m_{\tt_1}$, $m_{\tt_2}$; $m_{\tb_1}$, $m_{\tb_2}$ respectively, and their
 mixing angles $\theta_{\tt}$, $\theta_{\tb}$. \\
The $\tilde{\tau}$ mass matrix has a similar structure as the $\tb$ matrix,
eq.~(\ref{Msbottom}), with the replacements $M_Q \to M_L$,
$m_{\tb_R} \to m_{\tilde{\tau}_R}$, $m_b \to m_\tau$, 
and appropriate D-terms.

\section{De-diagonalization of M}\label{app2}
\setcounter{equation}{0}
The following discussion can be carried out in terms of any pair
of  $\tilde{M}_{N_i}$ (see section 3.2 for notations). 
We choose for definiteness
$ \tilde{M}_{N_1}, \tilde{M}_{N_4} $ and make the convenient
relabeling
$ \tilde{M}_{N_1} \equiv X, \tilde{M}_{N_4} \equiv Y$
and $ Z = X Y$. At this level we aim at the determination of some general 
consistency constraints and the distinction between  
input or output parameters is only for convenience.\\
 
Using Eq.(\ref{inv4}) to substitute for $Z$
in Eqs.(\ref{inv1} - \ref{inv3}), one obtains a set of three
linear equations in, say, 
the variables 
$X, Y, M_1$. This system
would give easily  $X, Y$ and  $M_1$ in terms of the other parameters
in the form

\begin{equation}
X = \frac{\Delta_X}{\Delta}, \,\,Y = \frac{\Delta_Y}{\Delta}, 
\,\, M_1 = \frac{\Delta_{M_1}}{\Delta}
\end{equation}

where it not for the fact that generically $\Delta = 0$ while
$\Delta_X \neq 0$ and $\Delta_Y \neq 0$. Actually the same happens
independently for the two subsystems Eqs.(\ref{inv1}, \ref{inv2})
or Eqs.(\ref{inv1}, \ref{inv3}) when one tries to solve
for $X, Y$, leading respectively to
\begin{equation}
X = \frac{\delta_1}{\delta} \sim Y  
\end{equation}

and
\begin{equation}
X = \frac{\delta_2}{\delta} \sim Y  
\end{equation}

where $\delta = 0$ and $\delta_1$, $\delta_2$  are generically non vanishing. 
The necessary and sufficient conditions for the existence 
of solutions are thus $\delta_1 = 0$ and $\delta_2 = 0$. They can be
cast in the following form:

\begin{eqnarray}
&&  P_{i j}^2  
+  (\mu^2 + M_Z^2 - M_1 M_2 + (M_1 + M_2) S_{i j} - S_{i j}^2) P_{i j} 
\nonumber \\
&& + \mu M_Z^2 (c_w^2 M_1 + s_w^2 M_2 ) \sin 2 \beta -\mu^2 M_1 M_2 =0
\label{condition1} \\
\nonumber
\end{eqnarray}

and

\begin{eqnarray}
&&  (M_1 + M_2 - S_{i j}) P_{i j}^2+  (\mu^2 (M_1 + M_2)  + 
            M_Z^2 (c_w^2 M_1 + s_w^2 M_2  - \mu \sin 2 \beta)) P_{i j} 
\nonumber \\
&&+ \mu ( M_Z^2 (c_w^2 M_1 + s_w^2 M_2 ) \sin 2 \beta -\mu M_1 M_2) 
    S_{i j}=0 
\label{condition2} \\
\nonumber
\end{eqnarray}

where $S_{i j}= \tilde{M}_{N_i} + \tilde{M}_{N_j}$ and
$ P_{i j}= \tilde{M}_{N_i} \tilde{M}_{N_j}$, $i=1,...,4, j=1,...,4$ 
with $i \neq j$.
Note that we used here the fact that equations
(\ref{condition1}) and (\ref{condition2}) should hold for any
pair of eigenvalues since we could have chosen as $(X, Y)$ any set 
$( \tilde{M}_{N_i}, \tilde{M}_{N_j})$ with $i \neq j$.

When Eq.(\ref{condition1}) is satisfied, Eqs.(\ref{inv1}) and (\ref{inv2}) 
become equivalent to each other, so that
the system made of (\ref{inv1}), (\ref{inv2}) and  (\ref{inv4}) is solvable
in terms
of $X$ and $Y$. Similarly, Eq.(\ref{condition2}) does the same for the system 
(\ref{inv1}) (\ref{inv3}) and (\ref{inv4}). The bottom line here is that
we have traded Eqs.(\ref{inv1} - \ref{inv4}) for the system  
(\ref{condition1}), (\ref{condition2}) which has, however, some welcome features
suitable for our purpose.  
An immediate consequence is that,
[within the strategy defined in section 3 where $M_2$ and $\mu$ are determined
from the physical chargino masses $M_{\chi^+_1}, M_{\chi^+_2}$ and $\tan \beta$], one can determine uniquely $M_1$ and any three neutralinos mass parameters,
say $\tilde{M}_{N_1}, \tilde{M}_{N_3}, \tilde{M}_{N_4}$ once
the fourth, say $\tilde{M}_{N_2}$ is given. Indeed, substituting $M_1$ from
Eq.(\ref{condition1}) into Eq.(\ref{condition2}) one obtains
the following cubic equation in $\tilde{M}_{N_i}, i\neq 2$,

\begin{equation}
a_1 \tilde{M}_{N_i}^3 + a_2 \tilde{M}_{N_i}^2 + a_3 \tilde{M}_{N_i} + a_4 =0 \label{cubiceq}
\end{equation}

where

\begin{eqnarray}
&a_1&= \tilde{M}_{N_2}^3 + M_2 ( \mu^2 -\tilde{M}_{N_2}^2) 
 - \tilde{M}_{N_2} (\mu^2 + c_w^2 M_Z^2) - 
     c_w^2 \mu M_Z^2 \sin 2\beta \nonumber \\
&a_2&= s_w^2 M_Z^2 ( \tilde{M}_{N_2} - M_2) (\tilde{M}_{N_2} + \mu
\sin 2 \beta) - M_2 a_1 \nonumber \\
&a_3&= s_w^2 M_Z^2 ( \tilde{M}_{N_2}^3 + (M_2 - \tilde{M}_{N_2})^2 
\mu \sin 2 \beta + M_2 \tilde{M}_{N_2} (M_2 - 2 \tilde{M}_{N_2}))
- ( \mu^2 + M_Z^2) a_1  \nonumber \\
&a_4&= \mu [ s_w^2 M_Z^2 ( M_2 ( M_2 - \tilde{M}_{N_2}) 
(\mu + \tilde{M}_{N_2} \sin 2 \beta) + c_w^2 M_Z^2 \sin 2\beta (
\mu \sin 2 \beta + \tilde{M}_{N_2}) )  
\nonumber \\
&& \,\,\,\,\,\,\,\,\,\,\,\,\,\,\,\,
+(\mu M_2 - c_w^2 M_Z^2 \sin 2 \beta) a_1 ] \nonumber \\
\label{ais}
\end{eqnarray}
 
The three solutions of Eq.(\ref{cubiceq})
are given by
 
\begin{equation}
\tilde{M}_{N_1}=
\frac{1}{3 a_1}(-a_2 - \frac{\sqrt[3]{2} A}{B} + \frac{B}{\sqrt[3]{2}})
\label{sol1}
\end{equation}

\begin{equation}
\tilde{M}_{N_3}=
\frac{1}{3 a_1}(-a_2 + \frac{(1 + I \sqrt{3}) A}{\sqrt[3]{4} B} - 
       \frac{(1 - I \sqrt{3}) B}{2 \sqrt[3]{2}}) 
\label{sol2}
\end{equation}
\begin{equation}
\tilde{M}_{N_4}=
\frac{1}{3 a_1}(-a_2 + \frac{(1 - I \sqrt{3}) A}{\sqrt[3]{4} B} - 
       \frac{(1 + I \sqrt{3}) B}{2 \sqrt[3]{2}})
\label{sol3}
\end{equation}

with 

\begin{equation}
B=\sqrt[3]{C + \sqrt{4 A^3 + C^2}}
\label{eqB}
\end{equation}

\begin{equation}
A= -a_2^2 + 3 a_1 a_3
\label{eqA}
\end{equation}

\begin{equation}
C=-2 a_2^3 + 9 a_1 a_2 a_3 - 27 a_1^2 a_4
\label{eqC}
\end{equation}

These solutions correspond necessarily to the three unknown neutralino mass 
parameters $\tilde{M}_{N_1}, \tilde{M}_{N_3}, \tilde{M}_{N_4}$,
due to the fact that, as we mentioned before, 
Eqs.(\ref{condition1}, \ref{condition2}) should be satisfied for any
pair $(\tilde{M}_{N_i}, \tilde{M}_{N_j})$ with $i\neq j$.
Furthermore, injecting
any of these solutions back into Eq.(\ref{condition1}) determines the {\sl same}
and unique value for $M_1$ given in Eq.(\ref{solM1}).
Finally one can cast Eqs.(\ref{sol1}- \ref{sol3}) in an even simpler form
once it is realized that 
$\tilde{M}_{N_1}, \tilde{M}_{N_3}, \tilde{M}_{N_4}$ and $M_1$
are automatically real-valued
when $M_2, \mu$ (and $\tilde{M}_{N_2}$) are taken real. 
 
Indeed, when $\tilde{M}_{N_2}, M_2$ and $\mu$ are all real valued,
i.e. $a_1, a_2, a_3$ and $a_4$ real, 
the cubic equation (\ref{cubiceq}) assumes at least one real solution. 
The real-valuedness of $M_1$ follows then immediately from  
Eq.(\ref{solM1}). However, since the latter
equation should be satisfied for any pair 
$(\tilde{M}_{N_i}, \tilde{M}_{N_j})$ with $ i\neq j$, then all
the remaining neutralino mass eigenvalues must also be real.\\ 

It thus follows that one can rewrite Eqs.(\ref{sol1} - \ref{sol3})
in an explicitly real valued form. To do this, we note first that
the simultaneous reality of the $\tilde{M}_{N_i}$'s is equivalent
to

\begin{equation}
1 + \sqrt[3]{4}\frac{A}{|B|^2} = 0
\end{equation}
implying that A is necessarily a negative real-valued quantity.  
Plugging the above relation back into Eqs.(\ref{sol1} - \ref{sol3})
one finds Eqs.(\ref{solMN1} - \ref{solMN3})

\end{document}